\documentclass[nofootnotebib,pre,twocolumn,noshowpacs]{revtex4-1}
\usepackage{epsfig}
\usepackage[]{natbib}
\usepackage{xspace}
\usepackage{color}
\usepackage{bbold}
\usepackage{tabu}
\usepackage{mathtools}

\DeclarePairedDelimiter\bra{\langle}{|}
\DeclarePairedDelimiter\ket{|}{\rangle}

\begin{document}

\title{
Uncertainty relation between detection probability 
and  energy fluctuations  
 }

\author{Felix Thiel, Itay Mualem, David Kessler, Eli Barkai}
\affiliation{Department of Physics, Institute of Nanotechnology and Advanced Materials, Bar-Ilan University, Ramat-Gan
52900, Israel}

\begin{abstract}

A classical random  walker starting on a node of a finite graph 
will always reach any other node since the search is ergodic, namely it  is fully exploring space,
hence the arrival probability is unity.
For quantum walks, destructive interference may induce effectively non-ergodic features
in such search processes. 
We consider a tight-binding  quantum walk described by a time independent Hamiltonian $H$,
 starting in state $\ket{\psi_{{\rm in}}}$.
 Under repeated projective measurements,
 made on a target state $\ket{{\rm d}}$, the final detection
of the system is not guaranteed since destructive interference may split
the Hilbert space into a bright
subspace and an orthogonal dark one. Using this we find an uncertainty
relation for the deviations of the  detection probability $P_{{\rm det}}$ which reads
$$\Delta P 
\mbox{Var}(H)_d  
\ge  |\bra{{\rm d}} \left[ H, D \right] \ket{\psi_{{\rm in}}}|^2.$$ 
 Here 
$\Delta P= P_{{\rm det}} - |\bra{\psi_{{\rm in}}}{\rm d} \rangle|^2$ is the deviation
of the  total  detection probability from the initial probability of detection,
$\mbox{Var}(H)_d$ is the variance of the energy in the detected state, 
and on the right hand side of the inequality we use the commutation relation of $H$ and the measurement projector $D = \ket{{\rm d}} \bra{{\rm d}}$.
Extensions and examples are discussed.

% This is suggestion of Felix
% $\Delta P \mbox{Var}(H)_d   \ge  |\bra{{\rm d}} \comm*{H}{D} \ket{\psi_{{\rm in}}} |^2$. 

\end{abstract}
\maketitle

\section{Introduction}

  A classical random walk on a finite graph explores the system completely.
 In a search process a particle starts on some node of a graph and then searches for a target  on another node. 
For classical walks on such structures the arrival  probability $P_{{\rm det}}$, that is the probability that the particle
starting on one node and detected eleswhere,
in principle after some long time, is unity. 
In that sense classical random walk search on a finite graph is efficient though time wise the search can be very inefficient. 
The exceptions with $P_{\rm det}\neq 1$ are obvious.
If the graph is decomposed into several non-connected parts, the dynamics is not ergodic,
and then exploration of at least part of the  space is prohibeted.  
Thus, the related  recurrence problem 
becomes an issue only for an infinite system,
for example a classical random walk on a lattice  in dimension larger than two is non-recurrent
\cite{Polya,Redner,Ralf}. 

 A very different behavior  is found for quantum walkers \cite{Aharonov1,Ambainis,Blumen,Salvador}
 on finite graphs that start localized
on a node of the graph. Firstly the concept of quantum arrival is not well defined, and instead we discuss 
the first detection, see below. Secondly,   destructive interference may divide
the Hilbert space into two components called dark and bright, and this yields an effect similar
to classical non-ergodicity, $P_{{\rm det}}< 1$.  More  specifically,  an  observer performs
repeated strong  measurements, made on another  node, in an attempt to detect the particle \cite{Bach,Krovi,Krovi1,Varbanov,Grunbaum,Krapivsky,Dhar,Dhar1,Sinkovicz1,Harel,Felix1,FelixPRA,Lahiri,Dubey,Liu}. In the time intervals between the measurements the dynamics is unitary. 
 The rate of measurement attempts is $1/\tau$ where
$\tau$ is a parameter of choice (see details below). 
 Due to destructive interference, there might
 exist certain initial states whose  amplitude  vanishes at the detected
node at all times  and this renders them non-detectable. 
Such initial conditions are called dark states and they are
widely encountered \cite{Caruso}.
In this case the mean hitting time, i.e. the mean time for detection, 
is infinite \cite{Krovi1}  and this can be found for simple models like
a quantum walk on a hyper-cube \cite{Krovi}  or a ring \cite{Harel}.
 More precisely to get non-classical behavior
for $P_{{\rm det}}$ 
 the system has to have some 
symmetry built into it \cite{Krovi1,FelixLET}. 
 Generic initial states $\ket{\psi_{{\rm in}} }$  are  linear combinations
of dark and bright states, the latter are detected with probability one.
It follows that a system starting in state $\ket{\psi_{{\rm in}} } $ 
 has a probability to be detected that lies somewhere between
zero and unity. 
 The question remains is to quantify this 
probability $0\le P_{{\rm det}}\le 1$?

%\footnote{Submitted to: Entropy special issue, Axiomatic Approaches to Quantum Mechanics}

 A formal solution for the detection probability  was found in  \cite{Krovi,FelixLET}
and obtained explicitly for a few examples \cite{FelixExactKB}. 
 Since $P_{{\rm det}}$ is non-trivial if compared with the classical counterpart
 we will find its bound.
As announced in the abstract,
this  is  presented with an uncertainty relation, relating the detection
probability, fluctuations of energy and the commutation relation
of $H$ with the projection operator describing the measurement.   
For that aim we consider a measurement protocol with stroboscopic detection,
however we argue below that the results obtained here are generic.
Generally the uncertainty principle 
\cite{Busch}
describes the deviation of the quantum
world from classical  Newtonian mechanics. Our goal is very different. 
 We present an approach that
shows how  quantum walks depart from the corresponding 
classical walks.  
Note that the uncertainty principle discussed here,
an exact formal solution to the problem, and an upper bound based on symmetry were presented recently in a short communication  \cite{FelixLET}.

\section{Model and Notation}

 We consider quantum dynamics on a finite graph pierced by measurements. 
The evolution free of measurement is described by a time
independent Hamiltonian $H$ and the corresponding unitary propagator. 
Examples   
include a single particle on a graph, where $H$ is a tight-binding 
Hamiltonian,  e.g. dynamics of a particle on a  ring with hops to nearest neighbours. 
The theory is valid in generality e.g. by identifying the graph with
 a Fock space, one can describe the dynamics of a many body system,
see an example in Ref.  \cite{Yin}. 
Initially the particle is in state $\ket{\psi_{{\rm in}}}$ which could be
a state localized on a  node of the graph. 

We use stroboscopic measurements at times $\tau, 2 \tau, ... $ in an attempt
to detect the particle in state $\ket{{\rm d}}$, for example
on a node
of the graph or an energy eigen state of $H$ etc. 
Specifically
the measurement if successful
 projects the wave packet onto state $\ket{{\rm d}}$ otherwise this
state is projected out and the wave function renormalised (see below). 
Between the measurement attempts the evolution is unitary described
with $U= \exp( - i H \tau)$  and here $\hbar=1$. 
The outcome of a measurement is binary:  either a failure in detection (no) 
or success (yes). 
 Thus the string of
measurements yields a sequence no, no, and in the n-th attempt a yes, though the final
success is not a rule of nature. 
Once we detect the particle we are done, we say that $n \tau$ is the time
it took to detect the particle in state $\ket{{\rm d}}$. 
 Fixing $H$ and $\tau$ and repeating this measurement process does not mean that the particle is always detected.
The question addressed here is what is the probability that
the particle is detected at all, in principle
after an infinite number of measurement attempts. This is the total detection
probability $P_{{\rm det}}$. 

 This model is the quantum version of the first passage time
problem \cite{Redner,Ralf}. However, here the measurements backfire and modify the dynamics.
Specifically, each individual  measurement is described with the collapse postulate. 
Namely if the system's
wave function is $\ket{\psi}$ at the moment of the detection attempt, the amplitude of finding the particle at $\ket{{\rm d}}$ is $\bra{{\rm d}}\psi\rangle$. As mentioned if the particle
is detected we are done. If not the amplitude on the detector is reset to 
zero, the wave function renormalised, and the evolution free of measurements continues until the next
measurement etc. Mathematically the failed detection transforms $\ket{\psi}$,
i.e. the wave function just before measurement,  to $N( 1 - D)\ket{\psi}$ where
$N$ is the normalization, and $D =\ket{{\rm d} } \bra{{\rm d}}$ is the measurement projector.
As mentioned in the introduction, aspects of this first detection problem, were
investigated previously,  in several works.

\section{ Lower bound using the propagator}

A bright state is an initial condition that is eventually detected with probability one
$P_{{\rm det}} (\psi_{{\rm in}}) =1$, while a dark state is never detected hence $P_{{\rm det}} (\psi_{{\rm in}}) =0$.
  Following \cite{FelixLET,FelixExactKB} it is not difficult to show that 
if $\ket{ \beta}$ is a normalised bright state
then $U^{\dagger} \ket{\beta}$ is bright as well. 
 We will soon
present  simple arguments to 
explain this statement but first let us point out its usefulness.
Clearly it follows that 
 $(U^{\dagger})^k \ket{\beta}$ is bright  when  $k$ is a
 non-negative  integer. As a seed consider the following
state $\ket{\beta} = U^{\dagger} \ket{{\rm d}}$.
This is an obvious bright state since it is detected in the first measurement
attempt with probability one,
because $\bra{{\rm d}} U U^{\dagger} \ket{{\rm d}} = 1$. 
It therefore follows that the states
\begin{equation}
 U^{\dagger} \ket{{\rm d}} , \cdots , (U^{\dagger} )^k \ket{{\rm d}} \cdots 
\label{eqR1}
\end{equation}
are all bright. 

Why is $U^{\dagger} \ket{\beta}$ bright in the first place? As is well known, the energy basis 
is  complete. Less well known is 
that  we may construct a complete set of  stationary eigenstates of $H$
which are either  bright  
or dark \cite{FelixExactKB}. These form a complete set which a priori is
not trivial since in principle one could imagine a situation where
the Hilbert space is divided into (say) dark, bright, and grey states,
the latter are states  that are detected with probability less than one 
but larger
than zero. In \cite{FelixExactKB} we  
 present  a formula for the stationary dark and bright states, making this statement more explicit,
but for now all we need to know is that the finite Hilbert space
can be divided into these two orthogonal subspaces denoted ${\cal H}_B$ (bright)   and ${\cal H}_{D}$ (dark). 
 Let $\ket{E_{j} ^{D }}$ be a specific
stationary  dark
state, and $j$ is an index enumerating this family of  states. Then clearly
$\bra {E_{j} ^{D} } U^{\dagger} \ket{\beta} = 0$ since the dark state
is orthogonal to the bright one and  $\bra{E_{j} ^{D} } U^{\dagger}$ 
gives a phase $\exp( i E_j \tau)$ where $E_j$ is the corresponding energy level. It follows that $U^{\dagger} \ket{\beta}$ has no dark component
in it and hence it is bright. 
Thus the whole approach is based on the fact that we may divide a finite
Hilbert space to a dark and bright subspaces. These
are related to the so called Zeno sub-spaces
 used to constrain
dynamics to part of the Hilbert space \cite{Facchi,Pascazio,Gherardini}.
In contrast  we consider  generic initial conditions
that belong both to the dark and the bright components of the Hilbert space and
this gives non-trivial detection probability.

 We notice in the same way that the sequence 
\begin{equation}
U\ket{\beta}, \cdots U^k \ket{\beta} \cdots 
\label{eqR2}
\end{equation}
is also bright. As a consequence, since $U^{\dagger} \ket{{\rm d}}$ is bright,
the state $ U U^{\dagger} \ket{{\rm d}} $ is also bright.  So when starting
at the state
$\ket{{\rm d}}$ the system is detected with probability one,
as was shown previously using  other
methods \cite{Grunbaum}.   

 We now see that 
\begin{equation} 
 \ket{{\rm d}} , U^{\dagger} \ket{{\rm d}} \cdots 
\label{eqSeq}
\end{equation}
are bright states. However they are not orthogonal. We may choose the
first two terms and construct two bright 
orthogonal vectors
\begin{equation}
\ket{ \tilde{\chi}_1 } = \ket{{\rm d}} \ \mbox{and} \ 
 \ket{\tilde{\chi}_2}= N\left(\ket{{\rm d}} - { U^{\dagger} \ket{{\rm d}} \over \bra{{\rm d}} U^{\dagger} \ket{{\rm d}} } \right)
\label{eqtilCHI}
\end{equation}
where $N$ is a normalization constant and $\ket{{\rm d}}$, the detected state, 
 is not an eigenstate of $U^{\dagger}$. 
We note that using a similar approach 
it is not difficult 
to obtain further orthogonal bright states  though
in practice we perform this step (see below)
with a computer program to avoid the algebra. 
Further, the infinite  sequence in Eq. (\ref{eqSeq}) is over-complete. To construct a bright
space  we must find a set of orthogonal vectors forming a basis
using for example the Gram-Schmidt method.
Formally the bright space is
\begin{equation}
 {\cal H}_B = \mbox{Span} \{ \ket{{\rm d}}\ , U \ket{{\rm d}}\ , U^2 \ket{{\rm d}}\ ,\cdots \}
\label{eqSPAN}
\end{equation} 
where $\mbox{Span}\{ ... \}$ is the set containing all linear combinations of 
vectors of the states in the parentheses. 
Below we will construct this space explicitly for some simple examples, however
in general this demands crunching  linear algebra. For this reason the here-presented
 uncertainty  relation,
 is useful in many cases. 

When we have found a complete orthonormal  basis vectors, which are
 either dark or bright states,
the detection probability is given by
 \cite{Krovi};
\begin{equation}
 P_{{\rm det}} = \sum_{ \tilde{\beta} \in {\cal H}_B } |\bra{\tilde{\beta}} \psi_{{\rm in}}\rangle|^2,
\label{eqbright}
\end{equation}
where the summation is over a bright basis which as usual has many representations.
It follows that
\begin{equation}
 P_{{\rm det}} \ge | \bra{\tilde{\chi}_1} \psi_{{\rm in}} \rangle|^2  
+  | \bra{\tilde{\chi}_2} \psi_{{\rm in}}  \rangle|^2  . 
\label{eqchichi}
\end{equation}
We then find
\begin{equation}
 P_{{\rm det}} \ge { | \bra{{\rm d}} U \ket{\psi_{{\rm in}}} |^2 \over
1 - |\bra{{\rm d}} U \ket{{\rm d}} |^2} 
\end{equation}
provided that the initial state and the detected one
are orthogonal $\bra{{\rm d}} \psi_{{\rm in}}\rangle=\bra{\tilde{\chi}_1} \psi_{{\rm det}} \rangle =0$.
Note that $ U \ket{\psi_{{\rm in}}}$ and $ U \ket{{\rm d}} $ are solutions
of the Schr\"odinger equation in the absence of measurement starting
with $\ket{\psi_{{\rm in}}}$ and $\ket{{\rm d}}$ respectively,  hence the 
lower bound relates the dynamics of a measurement free process
at time $\tau$ to the detection probability which is the 
outcome of the repeated  measurement process. The bound generally depends
on $\tau$ and at least in principle one may search for
$\tau$ that maximizes its  right hand side.
When $\tau$ is small, we may expand the propagator
to second order $U\sim 1 - i H \tau - H^2 \tau^2/2$
and then find
\begin{equation}
 P_{{\rm det}} \ge { | \bra{{\rm d}} H \ket{\psi_{{\rm in}}} |^2 \over
\mbox{Var}(H)_{{\rm d}} } 
\label{eqBBB}
\end{equation}
where 
\begin{equation}
\mbox{Var}(H)_{{\rm d}} = \bra{{\rm d}} H^2 \ket{{\rm d}} - \bra{{\rm d}} H \ket{{\rm d}}^2 
\end{equation}
characterises  the fluctuation of the energy in
the detected state. Thus the detection probability is bounded
by the transition matrix from the  initial to the  detected  state
divided by the fluctuations of the energy in the latter.
What comes to us  as a surprise is that this result is valid for practically
any $\tau$,
as we will show after a few remarks. 

%{\bf  Remark 1:} In the derivation we excluded detection states which
%are stationary eigenstates of $H$, for which the variance of energy is
%zero. Specifically, $\ket{\chi_2}$ is not defined in this case. 

%{\bf Remark 2.} 
%Our process starts at time $t=0$ and the first measurement
%is performed at time  $\tau$ (protocol one). 
%Instead of that one can define  a measurement  process
%where the first measurement is performed at time $t=0$ (protocol two).
%Clearly the two processes yield the same results 
%once we shift the initial condition
%either with $U$ or $U^{\dagger}$ depending which protocol
%is the starting point.  
%The first
% protocol while starting with $U^{\dagger}\ket{{\rm d}}$ 
%is clearly equivalent to the second  process  which starts in state 
%$\ket{{\rm d}}$. In the second protocol this state is
% an obvious bright state
% since the system is trivially detected on the origin of time
%at the first
%measurement attempt at $t=0$. 
%We stick to protocol one simply because
%it seems more popular \cite{Grunbaum,Dhar,Dhar1,Harel}.  
%More importantly, the results obtained here are general
%since it is easy to shift initial condition to go from
%one protocol to another or to  other timings of the first
%detection.

{\bf Remark 1.} Our results are valid for finite size systems, like
finite graphs. For infinite systems, it is not always possible
to divide the Hilbert space into two sub-spaces dark and bright. For example
for a one-dimensional tight binding quantum walk on a lattice, with
jump amplitude to nearest neighbours, starting on a node called the
origin and measuring there, the non-zero
 detection probability is less than unity \cite{Harel}.
This means that $\ket{{\rm d}}$ is not bright for an infinite system, which is
 physically  obvious as the wave packet can spread to infinity and hence the 
particle can escape detection. In this case $\ket{{\rm d}}$ is a grey state.

{\bf Remark 2.} The states $\ket{{\rm d}}, ... U^k \ket{{\rm d}} ... $ are bright
and similarly with $U^{\dagger}$.
Can we find a bright state $\ket{\beta'}$ orthogonal to these states?
The answer is negative, and hence these states can be used to span
the bright subspace. Assume $\ket{\beta'}$ is an initial condition
$\ket{\psi_{{\rm in}}}$ which is bright. At first measurement at time
$\tau$ the amplitude of detection is $\bra{{\rm d}} U \ket{\beta'}$ however
this is zero by the assumption that $\ket{\beta'}$ is orthogonal to the
just obtained set of bright states. We may continue with this
reasoning for the second, third, etc measurements, and we see that
amplitude of detection of $\ket{\beta'}$ is always zero.
It follows that state $\ket{\beta'}$ is not bright.

\section{Uncertainty relation}

 Let $\ket{\beta}$ be a bright state, then also $f(H)\ket{\beta}$ is bright
where $f(.)$ is an analytical function. 
Indeed similar to the previous section $\bra{E_{j} ^{D}} f(H) \ket{\beta}=0$ and
hence the state $f(H) \ket{\beta}$ has no dark element,  meaning it is bright.
As we showed already the state $\ket{{\rm d}}$ is bright so we
find a sequence of bright states
\begin{equation}
 \ket{{\rm d}}, H \ket{{\rm d}} , \cdots, H^k \ket{{\rm d}} \cdots. 
\label{eq09}
\end{equation}
We use the same approach  as in the previous section, namely we
use the first two states  and find two orthonormal bright states
\begin{equation}
 \ket{\chi_1} = \ket{{\rm d}}  \ \mbox{and} \
\ket{\chi_2} = N \left( \ket{{\rm d}} - { H \ket{{\rm d}} \over 
\bra{{\rm d}} H \ket{{\rm d}} } \right). 
\label{eq109}
\end{equation}
The normalization constant is given by $|N|^2 =  (\langle  H\rangle_{\rm d})^2 /\mbox{Var} (H)_{\rm d}$ 
where $\langle H\rangle_{\rm d} = \bra{{\rm d}} H \ket{{\rm d}}$. 
Since $H$ is Hermitian, inserting Eq. (\ref{eq109}) in 
Eq. (\ref{eqchichi}),  
and assuming no overlap of the initial
state with the detection one, $\bra{{\rm d}} \psi_{{\rm in}} \rangle=0$
we get 
Eq. (\ref{eqBBB}). 
Thus that  formula 
is valid for any $\tau$.

 For the more general case when the initial overlap with the
detection state is not zero, we define
\begin{equation}
 \Delta P = P_{{\rm det}} - | \bra{{\rm d}} \psi_{{\rm in}} \rangle|^2. 
\end{equation}
Since
$| \bra{{\rm d}} \psi_{{\rm in}} \rangle|^2$ is the square of
the  overlap of the initial state and the detected one,
 it gives the probability to detect the particle in a single-shot
 measurement
at time $t=0$. So $\Delta P$ is the difference between the
probability of detection after repeated measurements and the initial
probability of detection. Using Eqs.
(\ref{eqchichi},\ref{eq109})
and $
  \bra{{\rm d}}  H(D -1)  \ket{\psi_{{\rm in}}} =
  \bra{{\rm d}} \left[ H, D \right] \ket{\psi_{{\rm in}}}$ 
we find
\begin{equation}
\Delta P \mbox{ Var}(H)_{{\rm d}}  \ge | \bra{{\rm d}} \left[ H,D \right] \ket{\psi_{{\rm in}}}|^2.
\label{eqRob}
\end{equation}
Here $\left[H, D  \right] $ is the commutator of the Hamiltonian and the projector describing
the measurement.  
If $\ket{\psi_{{\rm in}}}= \ket{{\rm d}}$ the right hand side is equal zero and we find
$P_{{\rm det}} \ge 1$, hence here the
uncertainty relation indicates that the detection probability is unity. 
%

%{\bf Remark 5. } The relation Eq. (\ref{eqUNC})
% shows that $\Delta P$ is never negative, which is expected since repeated  measurements are always
%doing better than a single one.
% The pundit will notice that even this is not
%completely  obvious since the reference measurement is at the origin
%of time, where in practice we do not perform any measurement. 

{\bf Remark 3.} The matrix element $\bra{{\rm d}} H \ket{{\rm d}}$ can always  be set to be non-zero by a global shift
of the energy, hence $\ket{{\rm \chi}_2}$ is well defined. In the final result we can switch back to
any choice of energy scale, since the $\mbox{Var}(H)_d$  is  insensitive to the definition
of the zero of energy. 

{\bf Remark 4.} For the stroboscopic sampling, recurrences and revivals imply that
special sampling times $\tau$ defined through  $\Delta E \tau = 2 \pi k$ 
exhibit resonances \cite{Grunbaum,Harel} 
 such that the bounds based on energy are invalid. 
 Here $\Delta E$ is the energy difference between
any pair of energy levels in the system. 
In this case the starting point of the analysis should be 
 Eq. (\ref{eqR1})
and not Eq. 
(\ref{eq09}).

{\bf Remark 5.} We believe that our results are generally
valid for other detection protocols, for examples when we sample the system
randomly in time following a Poissonian process \cite{Varbanov}. 
This is because the dark and bright spaces are not sensitive
to the timing of the measurements.
Exceptional $\tau$'s mentioned in the previous
remark are clearly not something to worry about
in this more general case.

\section{The reverse dark approach}

Assume $\ket{\delta}$ is a normalised dark state, then as before 
the states $U^k \ket{\delta}, H^k \ket{\delta}, f(H) \ket{\delta}$ are dark as well. In fact
since by definition a system initially in  a dark state 
$\ket{\psi_{{\rm in }}} = \ket{\delta}$ is never detected,  it follows 
immediately that  $U^k \ket{\delta}$ and $H^k\ket{\delta}$ are dark (see
remark below).  However, there is 
no symmetry between dark and bright states, in the sense that, every system has
at least one bright state
$\ket{{\rm d}}$, but not every system necessarily has a dark state. 
 As we show in
\cite{FelixExactKB}
totally bright systems have non-degenerate energy levels
 and  all the energy eigenstates having a finite 
overlap
with the detected state. 
Still let us assume that we find a state $\ket{\delta}$, which
is dark, 
we can then apply nearly the same strategy as before to find an upper bound
for $P_{{\rm det}}$. We use $\ket{\delta}$ and
$H \ket{\delta}$ to construct two orthonormal dark states 
\begin{equation}
 \ket{\xi_1} = \ket{\delta}, \ \   \ket{\xi_2} = N\left( \ket{\delta} -{ H \ket{\delta} \over 
\bra{\delta} H \ket{\delta}  } \right) ,  
\end{equation}
where $N$ is a normalization constant.
Clearly here we assume that the dark state $\ket{\delta}$ is not a stationary 
state of the
system, since otherwise $\ket{\xi_2}$ is not defined. 
Analogous to Eq. (\ref{eqbright})
the detection probability is \cite{FelixExactKB}
\begin{equation}
P_{{\rm det}} = 1 - \sum_{\tilde{\delta} \in {\cal H}_{D}  }  |\bra{\tilde{\delta}} \psi_{{\rm in}} \rangle|^2. 
\label{eqSM}
\end{equation}
Here the summation is over a basis of the  subspace ${\cal H}_{D}$. 
 Since all the terms in the sum are clearly non-negative  
$$ P_{{\rm det}} \le 1- |\bra{\xi_1} \psi_{{\rm in}}\rangle|^2 -
 |\bra{\xi_2} \psi_{{\rm in}}\rangle|^2. $$
For simplicity assume that $\bra{\delta} \psi_{{\rm in}} \rangle=0$ then
using the normalization $N$ we find
\begin{equation}
 P_{{\rm det} } \le 1-  { |\bra{\delta} H \ket{ \psi_{{\rm in}}}|^2  \over \mbox{Var}(H)_\delta }. 
\end{equation}
Now we have an upper bound. As mentioned in the introduction, 
the detection probability even for small quantum systems on a graph,
can be less than unity, unlike the corresponding classical walks. 
So this is a useful bound provided that we can identify $\ket{\delta}$.
Notice that the variance of energy is now obtained with respect to the
dark state $\ket{\delta}$. Of course this state is dark with respect to a
state $\ket{{\rm d}}$ so while the detector does not appear explicitly
in our formula, it is obviously important.

We now relax the condition $\bra{\delta} \psi_{{\rm in}} \rangle=0$.
 Let 
 $P_{{\small ND}}= 1- |\bra{\delta} \psi_{{\rm in}}\rangle|^2$ be the probability that initially the
system is not in the dark state $\ket{\delta}$
(hence the subscript ${\small ND}$).
 We consider the deviation $\delta P= P_{ND}-P_{{\rm det}}$,  and find
\begin{equation}
 \delta P \mbox{Var}(H)_\delta \ge | \bra{\delta} \left[H , \ket{\delta} \bra{\delta} \right] \ket{\psi_{{\rm in}} } |^2. 
\label{eqUUPP}
\end{equation}
Unlike the lower bound 
Eq. (\ref{eqRob}),
here the commutator on the right hand side,
is between the  Hamiltonian and a projector of the 
dark state while previously the commutator was of $H$ and
the operator describing the measurement.

%{\bf Remark 9.} If $\ket{\delta}$ is an  eigenstate of $H$, we have 
%$\mbox{Var}(H)_\delta=0$ and also 
%$| \bra{\delta} H \hat{P}_\delta \ket{\psi_{{\rm in}} } |^2=0$ so the uncertainty relation
%gives $\delta P \times 0 \ge 0$ which is correct but hardly informative.
%As mentioned for this case $\ket{\xi_2}=0$. 
%Then we have the trivial
%bound  $P_{{\rm det}}\le 1 - |\bra{\delta} \psi_{{\rm in}}\rangle|^2$
%and if $\ket{\delta}$ is the only dark state then we get an equality
%(such an example will follow). 

%{\bf Remark 11.} We say that a state $\ket{\delta}$ 
% is globally dark if
%$\bra{{\rm d}} \exp( - i H \tau) \ket{\delta} =0$ for any $\tau$. 
%Expanding $0= \sum_{k=0} ^\infty\bra{{\rm d}}  (- i H \tau))^k \ket{\delta}$ and
%since $\tau$ is arbitrary $\bra{{\rm d}} H^k \ket{\delta}=0$ and hence
%$H^k \ket{\delta}$ is dark. On the other hand we may define a dark state
%with respect to the stroboscopic sampling only (which is far less general)
%and then   
%$\bra{{\rm d}} \exp( - i H k \tau) \ket{\delta} =0$ for specific value of
% $\tau$  and any non negative $k$. Such state is dark with respect to 
%stroboscopic measurements only, and here $U^k\ket{\delta}$ is dark 
%where $U=exp(- i H \tau)$, 

\section{Further improvement of the uncertainty relation}

 When $\bra{{\rm d}} [H, D  \ket{\psi_{{\rm in}}}=0$ the uncertainty  relation Eq.
(\ref{eqRob}) does not provide useful  information on $\Delta P$. 
Such situations can be found
for quantum walks on a graph and we encounter them frequently in the 
example section.  
This effect is easy to understand.   Assume that initially the particle is localised
on a node of a graph  while it is detected on a another distant
 node, in particular $\bra{{\rm d}} \psi_{{\rm in}} \rangle= 0$.
Then if  the hopping amplitudes are short ranged, 
the matrix element of $H$ between the two  distant nodes is  zero and  then 
the lower bound is not useful. 

For such cases we consider two other  orthonormal  bright states
$$ \ket{\chi_1} = \ket{{\rm d}} \  \mbox{and} \ \ket{\chi_2} = N\left( \ket{{\rm d}} - {H^s \ket{{\rm d}} \over \bra{{\rm d}} H^s \ket{{\rm d}} }\right) $$
where $s$ is 
a positive integer and  when $s=1$ we get the previously examined case. 
The normalization is $|N|^2= (\langle H^s \rangle_d)^2/ \mbox{Var} (H^s)_d$
where as before
$\langle H^s \rangle_d = \bra{{\rm d}} H^s \ket{{\rm d}}$ and 
$\mbox{Var} ( H^s) \rangle_d =  \bra{{\rm d}} H^{2 s} \ket{{\rm d}} - \bra{{\rm d}} H^s \ket{{\rm d}}^2$. 
Using Eq.
(\ref{eqchichi})
 we  find
\begin{equation}
\Delta P 
\mbox{Var}(H^s)_d  
\ge  | \bra{{\rm d}} \left[H^s ,D  \right]  \ket{\psi_{{\rm in}}} |^2.
\label{eqs}
\end{equation}
This relation between $\Delta P$, 
and the commutator of $H^s$ and the projector $D$ 
is clearly $s$ dependent. 
We will follow two approaches. 
  In the first we choose the smallest  $s$ such that the right hand side of the uncertainty relation
is not equal zero.  
The second  is to  choose $s$ in such a way to maximize the lower bound of
$P_{{\rm det}}$.
In the first approach and if $H$ is described by an adjacency matrix, $s$ has a simple physical
meaning as it is the distance between
the assumed localised  initial state and the detector state,
see further details below. 
The  first approach is the quickest way to gain insight, while the
second can be used to systematically  improve the result. 

\begin{figure}
\centering
\includegraphics[width=0.99\columnwidth]{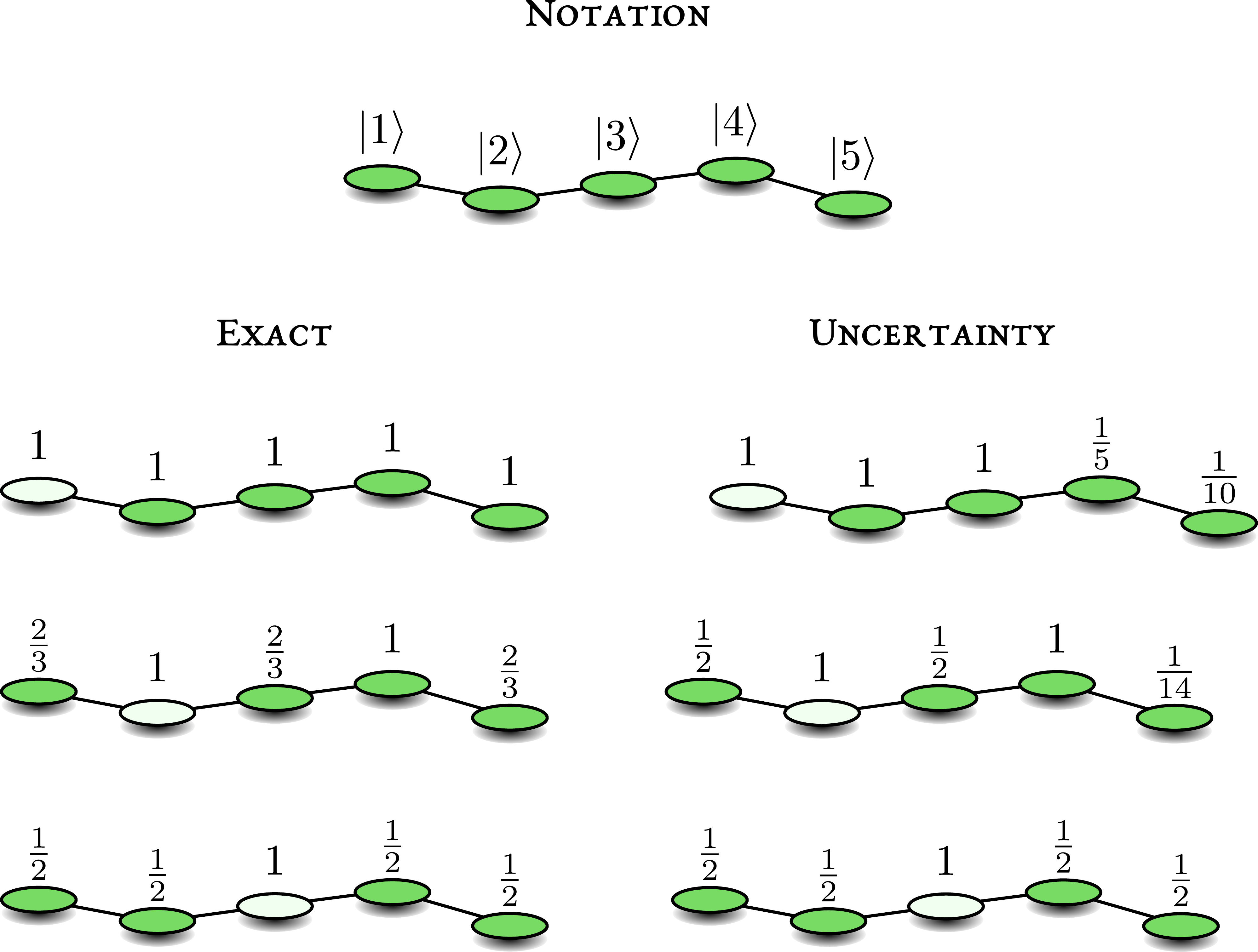}
\caption{
A quantum walk on a line with five sites. The initial condition is any
localised
state denoted with circles with an interior coloured green. 
  The detected state $\ket{{\rm d}}$ is the circle with a white  interior.  
Here we present the notation used in the text, the exact results of $P_{{\rm det}}$ and the lower bound found using the uncertainty relation. 
For example starting on $\ket{1}$ and measuring on $\ket{2}$ we have $P_{{\rm det}} =2/3$ while the uncertainty relation
gives
$P_{{\rm det}} \ge  1/2$.  
}
      \label{figSeg}
    \end{figure}

\section{Quantum walks on graphs}

We consider a quantum walk of a  single particle on a graph modelled with a  
 tight-binding Hamiltonian. 
In our examples $H$ is described by an adjacency matrix.
Thus the particle can occupy nodes of a graph, the edges/connections describe
hopping amplitudes. All these amplitudes are identical and on site energies
are set to zero. In the schematic figures of graphs under investigation, the
circle with the light interior  on a vertex describes the measured state $\ket{{\rm d}}$, e.g. Fig. 
\ref{figSeg}. We will
focus on initial conditions localised on another node (full circle) and
also on the  initial condition spread uniformly on the graph. 
Our goal is to find exact expressions for $P_{\rm det}$ using simple examples and compare the latter to the uncertainty relation. 

\subsection{Finite line}

 We consider a quantum walk on a finite line with $L$ nodes, 
 focusing on the example
of $L=5$. The localised basis is $\ket{r}$ with $r=1, \cdots L$,
so one may think of the end points as reflecting boundaries.
This will  soon
be compared
with a ring which has periodic boundaries. 
The Hamiltonian reads:
\begin{equation}
H = \gamma \left(
\begin{array}{ c c c c c} 
0 & 1 & 0 & 0 & 0  \\
1 & 0 & 1 & 0 & 0  \\
0 & 1 & 0 & 1 & 0  \\
0 & 0 & 1 & 0 & 1  \\
0 & 0 & 0 & 1 & 0  
\end{array}
\right).
\end{equation}
The hopping energy $\gamma$ is set to unity hereafter. 
Below we modify the location of the detector, and see its effect on 
the detection probability and the uncertainty principle. 
For schematics and summary of the  results see Fig. 
      \ref{figSeg}. 

{\em The transition to $\ket{{\rm d}} =\ket{1}$.}
Measurement is made on the left most node
$\ket{{\rm d}} = \ket{1}$ which from symmetry 
is the same as the choice $\ket{{\rm d}} = \ket{5}$. 
 We use $\ket{{\rm d}},H \ket{{\rm d}} \cdots $ to construct
 the bright subspace
  and besides obvious normalization we get 
%
%\begin{equation}
$$ {\cal H}_B = \mbox{Span}  \left\{\ket{1},\ket{2},\ket{1} + \ket{3} ,2 \ket{2} + \ket{4},
2 \ket{1} + 3 \ket{3} + \ket{5} \right\} . $$
%\end{equation}
%
Since the dimension of the Hilbert space is five and since the above 
states are linearly independent, we have no dark subspace. Hence
we are done:
the detection probability of any initial condition $\ket{\psi_{{\rm in}}}$
is unity. 

{\rm Aside.} For a quantum walk on a finite line, i.e. a lattice stretching
from $\ket{1}$ to $\ket{L}$, 
 with a time
independent $H$, and hops to nearest neighbours only (not necessarily translation invariant as in our example), 
 any initial state $\ket{\psi_{{\rm in}}}$ is detected
with probability one if the measurement is performed
on the end points  $\ket{1}$ or $\ket{L}$.   
This conclusion is immediate since the action of $H^k$ on
the detected state $\ket{1}$ gives $L$ linearly  independent vectors, which is
the dimension of the Hilbert space.

 Let us check the uncertainty principle starting with $\ket{\psi_{{\rm in}}}=\ket{2}$.
We have $\mbox{Var}(H)_1= \bra{1} H^2 \ket{2} - \bra{1} H \ket{1}^2 = 1$
and the transition amplitude $\bra{2} H \ket{1} =1$ hence
\begin{equation}
 P_{{\rm det}} (2\to 1) \ge { |\bra{2} H \ket{1} |^2 \over \mbox{Var}(H)_1} = 1. 
\end{equation}
So here the uncertainty principle gives the exact result, since clearly
$P_{{\rm det}} (2\to 1)\le 1$. 

  For the other starting points in the system, namely the transitions
$\ket{r} \to \ket{1}$ we use $s=r-1$ namely $s$ is  equal to the distance between the initial
state and detected one.  Our results are summarized in Fig. 
      \ref{figSeg}
and read: 
$P_{{\rm det}} (2 \to 1) \ge 1, P_{{\rm det }} (3 \to 1) \ge 1,
P_{{\rm det}}(4 \to 1)  \ge 1/5, P_{{\rm det}} (5 \to 1)\ge  1/10$.
We see that more distant initial states 
 depart from the exact result $P_{{\rm det}}=1$.
If $\ket{\psi_{{\rm in}}}= \ket{{\rm d}}$ we mentioned already that
 the uncertainty relation with $s=0$ 
gives $P_{{\rm det}} \ge 1$. This means that the particle is detected
with probability one $P_{{\rm det}}(1\to 1)=1$. 

So far we had no
overlap between initial and detected states.  For the uniform initial state
$\ket{\psi_{{\rm in}} } = \sum_{r=1} ^5 \ket{r}/\sqrt{5}$ we find
$P_{{\rm det}} \ge 6/25$ where we used $s=1$ (as mentioned the exact
detection probability is unity).
We will later optimise the choice of $s$ to see how one may improve the
prediction of the uncertainty relation.

{\em Detector on $\ket{{\rm d}}=\ket{2}$.} The bright subspace is spanned by
$\ket{{\rm d}}, H \ket{{\rm d}},  \cdots H^3 \ket{{\rm d}}$ namely
%
%\begin{equation}
$$ {\cal H}_B =\mbox{Span} \left\{   \ket{2}\ , \ket{1} +\ket{3}\ , 2 \ket{2} + \ket{4}\ , 2 \ket{1} + 3 \ket{3} + \ket{5} \right\}. $$
%\end{equation}
%
Here $H^4 \ket{{\rm d}}$ is also bright but is easily shown to be a linear
combination of $\ket{{\rm d}}$ and $H^2 \ket{{\rm d}}$. From these states
it is easy  construct an orthonormal bright basis
\begin{equation}
  \left\{ \ket{2}\ , (\ket{1} + \ket{3})/\sqrt{2}\ , \ket{4}\ , (2 \ket{5} + \ket{3} - \ket{1})/\sqrt{6}\right\}.
\label{eqBasis}
\end{equation}
 Hence the dimension
of the bright subspace is four, and the dark subspace has one
state in it. This vector is orthogonal to the bright basis Eq, (\ref{eqBasis}) 
 and
hence it  is easy to see that it is given by
\begin{equation}
\ket{\delta} = { \ket{1}  - \ket{3} + \ket{5} \over \sqrt{3} }.
\end{equation}
We have $H \ket{\delta}=0$; hence this state is a stationary state with an
 eigenvalue 
equal to zero. It is easy to see why this is a dark state: its amplitude
 on the detector 
is zero for any time.  Using 
Eq. (\ref{eqSM})
$P_{{\rm det}} = 1 - |\bra{\delta} \psi_{{\rm in}} \rangle |^2$ we get the exact values 
of
$P_{{\rm det}}(1 \to 2) = P_{{\rm det }} (3 \to 2) = P_{{\rm det}} (5 \to 2)=2/3$
and 
$P_{{\rm det}} (4 \to 2)=P_{{\rm det }} (2 \to 2)=1$.   
Compared with the case when the detector was on the edge we get values
for $P_{{\rm det}}$ which depart from the classical case of unit detection
probability. 
In Fig. 
      \ref{figSeg}.
we compare these results with the uncertainty principle:
$P_{{\rm det}} (1 \to 2) \ge 1/2, P_{{\rm det}} (3 \to 2) \ge 1/2, 
 P_{{\rm det}} (4 \to 2) \ge 1, P_{{\rm det}}(4\to 2)  \ge 1/14$,
where again we took $s$ to be the distance between initial and
detected state. 
For the uniform initial state we get $P_{{\rm det}} = 14/15$
while the uncertainty principle gives $P_{{\rm det}} \ge 13/25$
when we choose $s=1$. 

 It is easy to extend these results for a segment of length $L=2 k + 1$,
and $k=1,2, \cdots$. We measure on $\ket{2}$ or $\ket{L-1}$ 
and then  we have one and only one
dark state. This state is an eigenstate $\ket{\delta}=(1,0,-1, 0,1,0,-1,\cdots)/\sqrt{k+1}$ hence
\begin{equation}
P_{{\rm det}} (r \to 2) =
\left\{ 
\begin{array}{ c c }
{k \over k + 1}  & \ \mbox{if r is odd}  \\
1  & \ \mbox{ r is even} 
\end{array}
\right.
\end{equation}
On the other hand if $L$ is even we have no dark subspace. The detection
probability is unity. Thus depending  wether
$L$ is even or odd we may get a dark subspace or a completely bright situation.

{\em Detector on $\ket{{\rm d}}= \ket{3}$. } 
We now consider the detection on the middle point $\ket{3}$
as  this yields further symmetry in the problem. 
The  states $\ket{{\rm d}},H \ket{{\rm d}}$ and $H^2\ket{{\rm d}}$ are bright and
are easy to evaluate
$\left\{  \ket{3}\ , \ket{2} + \ket{4}\ , \ket{1} + 2 \ket{3} + \ket{5} \right\}.$
It is then easy to construct the dark space,
as we have only two dark states orthogonal to these bright states
a dark basis is $\left\{ \ket{\delta_1} , \ket{\delta_2}\right\}$ with
$\ket{\delta_1} =  (\ket{2} -\ket{4}) /\sqrt{2}$
and $\ket{\delta_2} = 
 (\ket {1} -\ket{5}) /\sqrt{2}$. 
 It is easy to understand why these states are
dark as they interfere destructively on the detector on $\ket{3}$ 
in such a way that the amplitude of detecting the particle there is zero.
We also have $H \ket{\delta_1} = \ket{\delta_2}$ and $H \ket{\delta_2} = \ket{\delta_1}$.
 Using the dark states and Eq. 
(\ref{eqSM})
we find that  $P_{{\rm det}} (r \to 3)=1/2$
for any initial state $\ket{r}$ the exception is the return problem
 $\ket{3} \to \ket{3}$ which is detected with probability one.  

  Considering the transition $\ket{2} \to \ket{3}$ we may use the two uncertainty relation, one with the detector state and the second with the dark state $\ket{\delta_2}$
to get:
\begin{equation}
 {1 \over 2} \le   P_{{\rm det}}(2 \to 3)  \le {1\over 2} . 
\end{equation}
So here the uncertainty relations give the exact result. 
The same holds for the transition $1 \to 3$ and the other transitions are
clearly identical from symmetry. The results are summarized in Fig. 
      \ref{figSeg}.

%\begin{widetext}
\begin{figure}
\centering
\includegraphics[width=0.99\columnwidth]{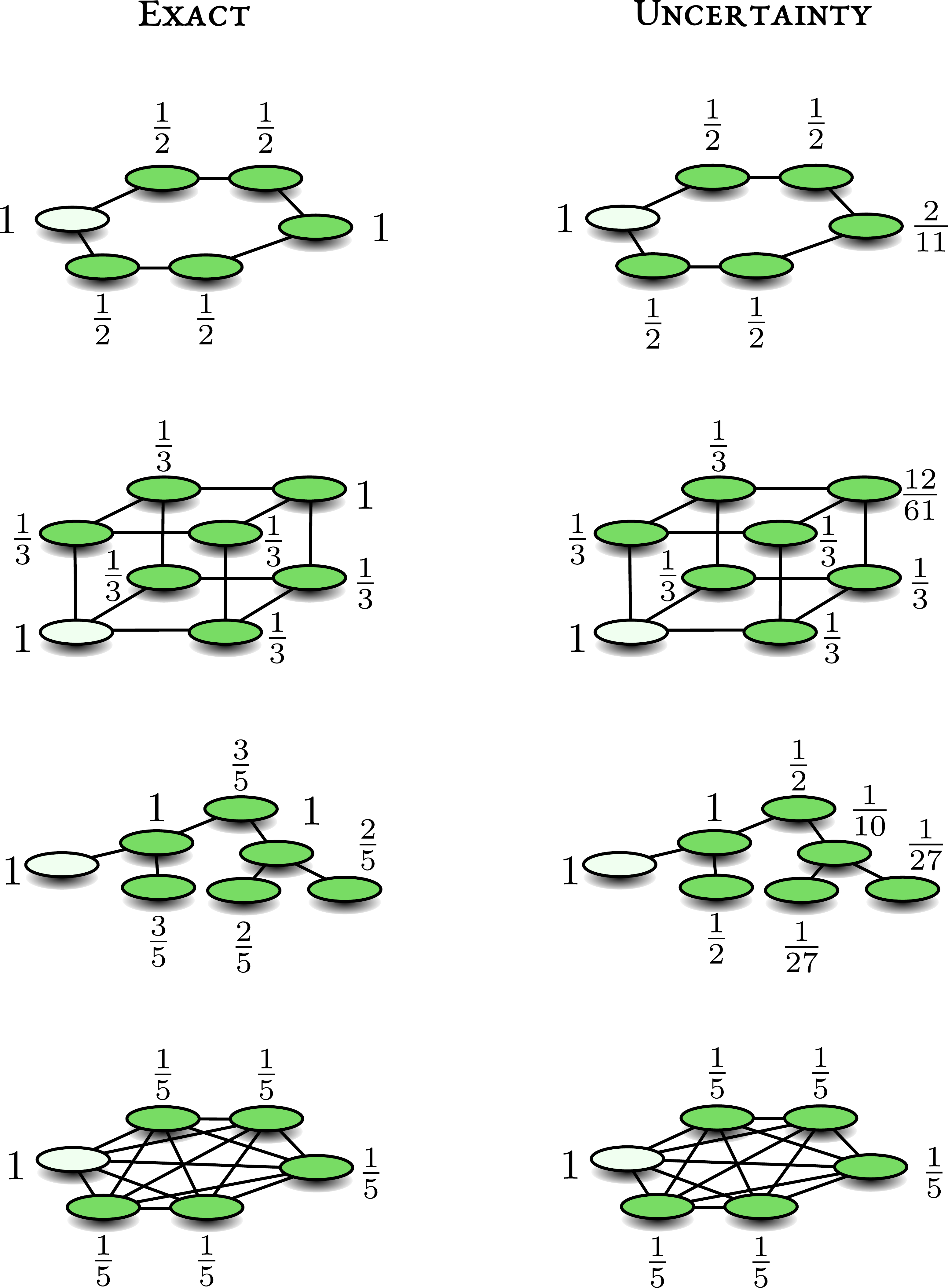}
\caption{
For a quantum walk starting on a node of the graph and measured elsewhere
we present the exact result for the detection probability which
depend's on the location of the  starting point. 
Also shown is the lower bound obtained with the uncertainty principle. We choose $s$ as the smallest integer for which the uncertainty principle is non trivial, namely the case
where $\bra{\psi_{{\rm in}}} H^s \ket{{\rm d}} \ne 0$, this is the distance between
the measured site and the initial condition.        }
      \label{figeexx}
    \end{figure}
%\end{widetext}

\subsection{Enumeration of paths approach}

 Let  $H=\gamma A$ where $A$ is the adjacency matrix of some graph. We set the energy
scale $\gamma=1$ as it does not control the detection probability (like the sampling rate $1/\tau$). 
Thus for the Hamiltonian under investigation,
 all bonds in the system are identical and on-site energies
are zero. So $H_{ii} =0$ while $H_{{ij}}=1$ if the states
are connected, zero otherwise. 
An example is  the segment of size
five just considered. As mentioned,  the detection
probability for the  return problem $\ket{{\rm d}} \to \ket{{\rm d}}$  
is unity, hence below we consider only the transition
problem from some localised initial state $\ket{r}$ to another orthogonal
localised state $\ket{{\rm d}}$.
In this case we can use a path counting approach to find a useful bound for
$P_{{\rm det}}$. 

We have
\begin{equation}
 H^s \ket{{\rm d}} = \sum_{\mbox{s paths}} \ket{j} 
\end{equation}
where the sum is over all states $\ket{j}$ which are end points of 
paths starting on $\ket{{\rm d}}$ and whose length is $s$. 
Clearly
\begin{equation}
 {\cal N}_{r \to d} (s) = \bra{r} H^s \ket{{\rm d}}
\end{equation}
is the  number of paths starting  on $\ket{r}$ and ending at
$\ket{{\rm d}}$ whose length is $s$. We then find from the uncertainty relation
\begin{equation}
P_{{\rm det}}(r \to {\rm d} ) \ge { \left[{\cal N}_{r \to {\rm d} } (s)\right]^2
\over {\cal N}_{{\rm d}  \to {\rm d} } (2 s) - \left[ {\cal N}_{{\rm d} \to {\rm d} } (s)\right]^2 }
\label{eq26}
\end{equation}
the denominator is the variance of $H^s$ in the detector state. 
For example when $\ket{r}$ is the nearest neighbour of $\ket{{\rm d}}$ we
choose $s=1$ and get 
\begin{equation}
 P_{{\rm det}} \left( r \to {\rm d}  \right) \ge { 1 \over \# n.n}
\end{equation}
where in the denominator we have the number of nearest neighbours. This is
the reason why for the example of the line we got for nearest neighbours
 $P_{{\rm det}} \ge 1/2$
while for the two  edge states, which have  only one nearest neighbour, we got
$P_{{\rm det}} \ge 1$. 
The bound depends on $s$ and this can be used to our advantage.
What is clear is that we must choose $s$ to be larger or equal to the distance
between the starting point to the measured one, otherwise the numerator
in Eq. (\ref{eq26}) is equal zero.

\subsection{The benzene like ring}

We now consider the tight-binding Hamiltonian for a particle on a ring with six sites  \cite{Harel}
\begin{equation}
 H =  \gamma \sum_{i=0} ^{5} [\ket{i} \bra{i+1} + \ket{i} \bra{i-1}  ]
\label{eqRING}
\end{equation}
see top left panel in  Fig. \ref{figeexx} for schematics.  
Here we use  periodic boundary conditions hence $\ket{6} =\ket{0}$. 
Similar to the previous example  the  energy scale $\gamma$
is irrelevant for the determination of 
$P_{{\rm det}}$ hence we may  set $\gamma=1$ and then $H$ is the adjacency matrix of the
benzene like  ring. 
To  see that $\gamma$ is irrelevant
note that both the left hand side and the right hand side of the uncertainty relation
Eq. 
(\ref{eqRob})
scale proportional to $\gamma$ hence it  can be canceled out. This holds
also from  the exact solution to the problem \cite{Harel} and is not limited to the example  under study \cite{FelixLET}.
 The repeated
measurements are made on a  site we call  $\ket{{\rm d}}=\ket{0}$ while from
symmetry all sites are identical. 

 With the method of enumeration of paths we consider three
transitions $1\to 0$, $2 \to 0$ and $3 \to 0$.
 Since the detector on the ring has two nearest neighbours, $P_{{\rm det}}(1 \to 0) \ge 1/2$. For the transition $2\to 0$ and 
$3 \to 0$ we will use $s=2$ and $s=3$ respectively. 
 Notice that we have only
one path of length two for the transition
$2 \to 0$  while for the transition $3 \to 0$ we have two paths
of length three. Elementary path counting gives
${\cal N}_{0\to 0 }(4)-  [{\cal N}_{0 \to 0} (2)]^2= 2$ for $s=2$
while  
${\cal N}_{0\to 0 }(6)-  [{\cal N}_{0 \to 0} (3)]^2= 22$ for $s=3$,
hence we get $P_{{\rm det}} (2\to 0)\ge 1/2$ and $P_{{\rm det}}(3 \to 0)\ge 2/11$.
These results are summarized in the upper right panel of
 Fig. \ref{figeexx}.  
We now analyse the problem exactly.

 In Eq. (\ref{eq09})
we showed that $\ket{\psi_{{\rm in}}}$ is proportional to any of the states
 $\ket{0}$, $H \ket{0}$, $H^2 \ket{0}$ and $H^3\ket{0}$ then it is bright. Also $H^4\ket{0}$ is bright, 
however to construct a basis
for the bright subspace ${\cal H}_B$  we need only the  
 just mentioned states.
Using $\ket{{\rm d}} = \ket{0}$ the bright subspace is given by
\begin{widetext}
$$ {\cal H}_B = \mbox{Span} \left\{ \ket{0}, {\ket{1} + \ket{5}\over \sqrt{2}} , {2 \ket{0} + \ket{2} + \ket{4} \over \sqrt{6} }, { 3 \ket{1} + 2 \ket{3} + 3 \ket{5} \over \sqrt{22} } \right\}.$$   
\end{widetext}
These states are nearly intuitively bright, for example the second
$(\ket{1} + \ket{5})/ \sqrt{2}$ is interfering constructively on the detector
situated 
on $\ket{0}$. 
On the other hand the state 
$(\ket{1} - \ket{5})/ \sqrt{2}$ is dark 
since from symmetry. The free evolution of this state gives zero amplitude
on the detector.  Acting with $H$ on this state and normalising
we see that $(\ket{2}-\ket{4})/\sqrt{2} $ is also dark which is again obvious
from symmetry. 

 With this information we can solve the problem exactly. For example,
consider a particle starting on 
\begin{equation}
\ket{1}={1 \over \sqrt{2}}  \left( 
\underbrace{{ \ket{1} + \ket{5 }\over \sqrt{2}}}_{\mbox{bright}}  + \underbrace{ {\ket{1} - \ket{5} \over \sqrt{2}}}_{\mbox{dark}}  \right). 
\end{equation}
As mentioned the first term  is bright and the second dark, hence
$P_{{\rm det}}(1\rightarrow  0) = 1/2$. Similarly $P_{{\rm det}}( 2\rightarrow
0) = 1/2$ and $P_{{\rm det }} (3\rightarrow 0) = 1$. 
The transition $5 \rightarrow 0$ is identical to $1 \rightarrow 0$
and similarly for $4\rightarrow 0$. 
 Let us now see how the lower and upper bounds work. 

{\em The $1 \rightarrow 0$ transition.}
As mentioned $\ket{{\rm d}} = \ket{0}$ and using Eq. 
(\ref{eqRING}) with $L-6$
 $\mbox{Var}(H)_0= 2$. For the initial condition $\ket{\psi_{{\rm in}}}=
\ket{1}$ we have zero initial overlap with the detector.
We find
\begin{equation}
 {1\over 2} = {|\bra{0} H \ket{1}|^2 \over \mbox{Var}(H)_0} \le 
P_{{\rm det}} \le 1 - { |\bra{\delta} H \ket{1}|^2 \over \mbox{Var}(H)_{\delta}}= {1 \over 2} 
\end{equation}
where  we used as a dark state $\ket{\delta} = (\ket{2} - \ket{4})/\sqrt{2}$ and
$\mbox{Var}(H)_{\delta}  = 1$. In this case the lower and upper bound coincide
giving the exact result $P_{{\rm det}} =1/2$.  

\begin{figure}
\centering
\includegraphics[width=0.99\columnwidth]{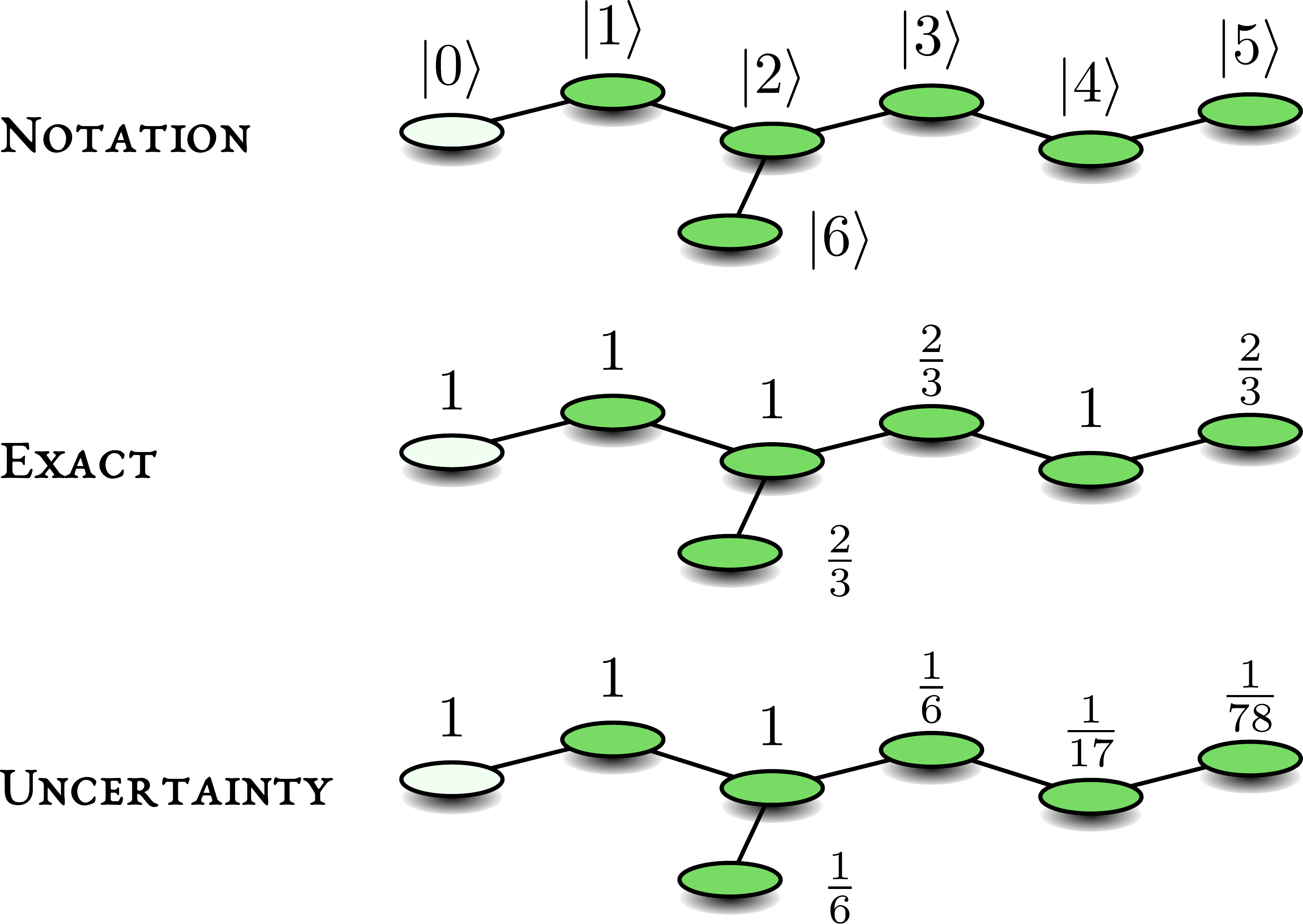}
\caption{
A system with a dangling bond where the edge state $\ket{0}$ is repeatedly
 measured.
Note that in the absence of the bond, i.e. when  node $\ket{6}$ is removed, 
 the detection probability is unity,
no matter what is the starting point.  
To obtain the lower bound using uncertainty we choose $s$ to be the shortest distance between the initial condition and the detected state. In Fig.
      \ref{figDang} we improve the bound for the transition 
$\ket{5} \to \ket{0}$, which here gives  $P_{{\rm det}} \ge 1/78$. 
}
      \label{figItai}
    \end{figure}

{\em The $2 \rightarrow 0$ transition.}
Since the matrix element $\bra{0} H \ket{2} =0$ we now 
use Eq. (\ref{eqs}) with $s=2$ since the distance between
the initial state and measured one is $2$. 
We find
\begin{equation}
 {1 \over 2} = {|\bra{2} H^2 \ket{0}|^2 \over \mbox{Var}(H^2)_{0}}
\le P_{{\rm det}} \le 1 - {\bra{\delta} H \ket{2}|^2 \over
 \mbox{Var}(H)_{\delta}  } = {1 \over 2} .
\end{equation}
 $\mbox{Var}(H^2)_0 = 2, \mbox{Var}(H)_{\delta} =1$ and $\ket{\delta} = (\ket{1} - \ket{5})/\sqrt{2}$. 
Again the lower and upper bound coincide giving the exact result.

{\em The $3 \rightarrow 0$ transition}. We now set $s=3$ finding $\mbox{Var}(H^3)_0 =22$, while $|\bra{0} H^3 \ket{3}|^2=4$ and this gives the mentioned
already  lower bound
$2/11 \le P_{{\rm det}} \le 1$ which is to compare with the exact
result  $P_{{\rm det}} =1$.

In general one
could imagine several ways of how to improve the lower  bound. 
The obvious one
is to go beyond the calculation of the bright states $\ket{{\rm d}}$
 and $H^s \ket{{\rm d}}$, namely consider a third bright state, 
but that means more algebra. 
Another option is to choose 
$H^{s_1} \ket{{\rm d}}$ and $H^{s_2} \ket{{\rm d}}$ as the starting point
(we took $s_1=0$ and $s_2=1,2,3$). In the next section we
will improve  the lower bound using a simple example. 
However we leave the optimisation
strategy of the lower bound for future work, as clearly the examples
we tackle here are  pedagogical.
The upper bound can be tackled with  symmetry arguments, which 
provide a different perspective \cite{FelixLET}.

\begin{figure}
\centering
\includegraphics[width=0.99\columnwidth]{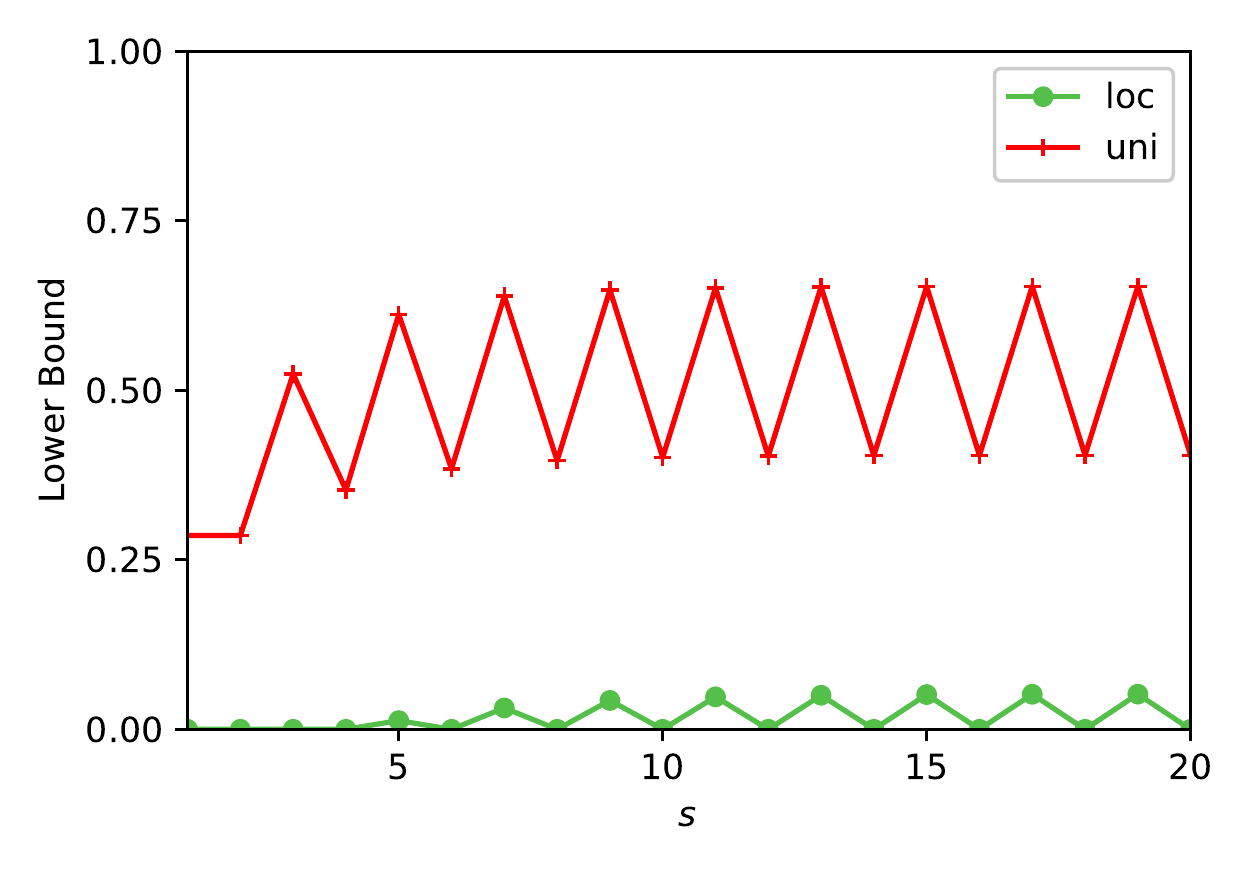}
\caption{
We demonstrate the 
optimisation of the  uncertainty principle for a system with a dangling bond.
The Fig. shows the lower bound for $P_{{\rm det}}$ versus $s$. The measurement is on
node $\ket{0}$ and we consider two initial conditions:
 one localised 
$\ket{\psi_{{\rm in}}} = \ket{5}$ and the other  uniform. See 
     upper part of Fig.  \ref{figItai} for schematics and notation.
 Starting on node $\ket{5}$ and when $s\le 4$ the bound is equal zero
since if the  initial and detected states are localised,  $s$  must be larger than the distance between these states 
to make the approach useful. 
Increasing $s$ clearly leads to an   improvements of the bound. 
The exact results are $P_{{\rm det}}= 20/21$ for the uniform initial condition,
and $P_{{\rm det}}= 2/3$ for the initially  localised state. 
}
      \label{figDang}
    \end{figure}

\subsection{Optimisations of the lower bound} 

Here we use a simple model of a system with a dangling bond, and
consider the optimisation of the lower bound, i.e. we will soon
search for the best choice of  $s$. 
 For the quantum walk on the line,  when 
the measurement is made on the end point, we showed that 
 the detection probability
is unity. We add a perturbation to the system: one
link perpendicular to the line.  We will treat the
example of  a system with $6$ nodes on
the backbone and one dangling bond, see schematics Fig.   
      \ref{figItai}.
Adding one dangling bond breaks the symmetry but only partially in the sense
that we still  find that the detection probability is not generically unity. We note that
strongly disordered
systems  with no symmetry exhibit a classical behaviour namely $P_{{\rm det}}=1$  \cite{FelixLET}. 

The notation used, the exact results, and the  uncertainty principle
are presented here in Fig. \ref{figItai} where the transition from
a localised initial state to the detector on $\ket{0}$ is considered. 
Here $s$ is the distance between the initial condition and the detected state.
Now we turn to a simple optimisation of the bound
both for a localised  initial state and for a uniform initial condition. 

 Consider the initial condition which is uniform
$\ket{\psi_{{\rm in}} } = \sum_{r=0} ^6\ket{r}/\sqrt{7}$.  
In this  example we have only  one dark state $\ket{\delta} = (\ket{6} - \ket{3} + \ket{5})/\sqrt{3}$, and with this we find the detection
probability $P_{{\rm det}}= 20/21$. The uncertainty principle, with $s=1$ 
reads:
\begin{equation}
 P_{{\rm det}} - {1 \over 7} \ge { | \bra{0} H \left( 1 - \ket{0} \bra{0}\right)\ket{ \psi_{{\rm in}} }|^2 \over \mbox{Var}(H)_0}= { 1 \over 7}  . 
\end{equation}
Here $1/7$ on the left hand side is the initial overlap,
and we used the detected node  $\ket{{\rm d}}=\ket{0}$. 
This gives $P_{{\rm det}} \ge 2/7$. To improve this bound we
made similar calculations with  $s=2,... 5$  the results are
 presented in Table $1$. 
The best lower bound is found for $s=5$, 
$P_{{\rm det}} \ge 167/253 \simeq 0.6117$.
To further improve the bound we depart from 
the § pen and paper approach instead using a simple 
program.  The results for both the uniform initial condition and
a localised initial state are presented in Fig. \ref{figDang}. 
This figure shows that increasing  $s$ yields a  better result for
the lower bound, but eventually the  calculation saturates while exhibiting odd/
even oscillation, 
Note that generally  the lower bound calculations
 involve only  the multiplication of matrices, 
while obtaining an orthonormal basis for the
dark ${\cal H}_D$ or bright ${\cal H}_B$ subspaces,  needed for the calculation of the exact
expression for $P_{{\rm det}}$ Eqs.
(\ref{eqbright},
\ref{eqSM}),
  demands considerably more work.

\begin{center}
\large
\begin{table} 
\begin{tabular}{|c | c |  c | c |  c | c| } 
\hline
$s$ & 1 & 2 & 3 & 4 & 5 \\
\hline
 $P_{{\rm det}} \ge$ \   & \  ${2 \over 7} $ & \  ${2 \over 7} $ & \  ${11 \over 21} $ & \  ${42 \over 119} $ &  \ ${167 \over 253} $  \\
\hline
\end{tabular}
\caption{For the system with a dangling bond, the detection being
on $\ket{0}$,
the Table gives
the lower bound for the uniform initial condition versus $s$.
Increasing $s$ in this range  is improving the bound.  
}
\end{table}
\end{center}

\subsection{Other examples}

 For not too large systems, 
 it is relatively easy to write a simple program  generating
the bright states $\ket{{\rm d}}, H \ket{{\rm d}}\cdots$
and  then to perform the Gram-Schmidt procedure.
This way we  find an  orthonormal basis for the bright space ${\cal H}_B$.
And then we can find  the detection probability using Eq. (\ref{eqbright}).
In Fig.  
      \ref{figeexx}
we present some graphs and the corresponding
detection probability.
Here as before  the starting point is a node of a graph, 
and the empty circle is the position of the detector.
It is striking, that when the detector is placed on
a symmetry centre of the graph, $P_{{\rm det}}$ may exhibit a deficit from
the classical limit of unity. 
Since the
classical detection probability on similar graphs is unity,
the deviations  are of interest.
So as a rule of thumb we use systems with  symmetry to find deviations 
from classical expectations, however
the symmetry may not be  always obvious, so it will be the subject
of our next paper.

 All the examples presented here are based on rather simple graphs.
This allowed us to obtain exact solutions rather easily  and then 
estimate the performance of the uncertainty principle. For large
and generally complex systems the bound  can perform rather poorly.
To tackle the problem one therefore needs a set of tools,
and one  cannot rely only on the method  presented here. We mention
briefly a few cases. If the system is disordered such that the energy spectrum
is non-degenerate, and each stationary state has some finite overlap with the
detected state, the detection probability is unity. In this case the
system behaves classically and there is no need at all for a lower bound.
For systems with symmetry, and hence a degenerate spectrum,
 an upper bound based on symmetry was found.
This relates the number of equivalent states in the system $\nu$, i.e.
states that are equivalent to the initial state with respect to the measured
one, showing that $P_{{\rm det}}\le 1/\nu$. Finally, here we used as a seed
the bright state $\ket{{\rm d}}$. In \cite{FelixLET}
we proposed a shell method, that searches for
a bright state, with a strong overlap with the detected one. This allowed
us to treat several large systems, like systems with loops, the hyper-cube,
and also go beyond the examples with adjacency Hamiltonians treated here
\cite{FelixLET}.

\section{Discussion}

 This work relied on the partition of the Hilbert space into
dark and bright sub-spaces. Probably the best known example, is the
case of rapid measurements, $\tau\to 0$ \cite{ZenoIN}. Then when measuring on
a node of the graph it is not difficult to find a  basis for these
sub-spaces. The measured node is obviously the bright sub-space,
since the particle is detected with probability one at the first measurement.
  All other nodes are dark, since the  probability of detection
is of order $\tau^2$ 
(unlike  classical walks where it increases like $\tau$) and hence 
$P_{\rm det}=0$ in the limit. It was later realised that the splitting of the Hilbert
space into two components, does not need to  rely on fast measurements
 \cite{Krovi1,Facchi,Pascazio,Gherardini}.
The general mechanism behind this effect is destructive interference.
Recently we expressed the dark and bright sub-space of a general Hamiltonian
in terms of its eigen-states 
\cite{FelixExactKB}.
Using this we could find here a bound for the
detection probability in the form of an uncertainty principle.  

The splitting of the Hilbert space into two components resembles the splitting
of a classical system into two disjoint components, namely ergodicity breaking.
Let us assume we have such a classical system, which is split into non-connected
 domains $A$ and $B$.
 We add a  detector within one sector  say $A$.
If we detect the particle then we know that it started in $A$ otherwise it
started in $B$. In the quantum world we may start in a superposition state,
with  components in both the dark and bright sub-spaces. Hence, this situation 
is very different if compared with the classical case of ergodicity breaking, 
leading to non-trivial
$P_{{\rm det}}$.

 The uncertainty   relation
Eq. 
(\ref{eqRob})
does not depend on the measurement frequency 
$1/\tau$ and in that sense it is universal. However in some
cases its right hand side is equal zero, in particular when an  initial
localised state and a detected localised state are far one from the other 
and the Hamiltonian
describes a finite range of jump amplitudes. 
 A second relation, 
Eq. (\ref{eqs})
depends on the free parameter $s$,  allowing to connect between distant states, and this permits 
an easy calculation of a nontrivial lower bound for $P_{{\rm det}}$. 
We showed how to optimise the choice of $s$ thus improving the lower
bound. 
More advanced methods are discussed in \cite{FelixLET}.

In this article we considered repeated strong measurement 
as the protocol of choice. Due to the wide range of quantum
 measurement theories
one must wonder how general are the results presented here? 
While the answer to this question is left for future work,
we may speculate the following. 
The mechanism leading to dark states is in principle  simple:
  the amplitude of the wave function at $\ket{{\rm d}}$ is  equal zero 
for ever.   Hence any choice of a measurement theory or any
measurement protocol,   
that is reasonably physical in the sense that
it postulates that we cannot detect neither influence
 the state of the  particle if the amplitude of finding it is zero,
will yield the same dark states as for strong measurements. 
Still we cannot claim any results for 
weak measurements \cite{Rozema}. 
We believe that our results hold also
for the well known non-Hermitian approach, where the detection is modelled with a sink.
In the limit of small $\tau$ it was shown \cite{Dhar,Lahiri} 
 that one may use a non-Hermitian approach to model the strong detection protocol considered here. 
Hence, the two approaches have many things in common. 
Instead of stroboscopic sampling one may use  temporal
 random sampling,  for example 
sampling times drawn from  a Poisson process \cite{Varbanov}. Again we believe
that this will not alter our results
since  the destructive interference is found also in this case. The fact that
our results are $\tau$ independent is another indication for the generality
of the approach. 

 The  uncertainty  principle
investigated here is different from the standard approaches \cite{Heisenberg,Kenard,Robertson,Vlad}. These are
roughly divided into two schools of thoughts.  
The text-book
momentum-position uncertainty relation, is a  measure
for uncertainty  in the state function. To verify it one needs
to perform two sets of measurements obtaining the uncertainty in
$x$ and $p$ independently. 
The second is the disturbance approach originating from the 
$\gamma$ ray thought experiment  \cite{Heisenberg}. 
This dichotomy has  attracted considerable  ongoing research until recently
\cite{Ozawa,Erhart,Oppenheim,Werner,STRONGER,ECOHEN}. 
Our approach is different from both and this is obviously related to the
fact that we consider repeated  measurements which
 backfire and  modify the unitary 
evolution and also to the observable of interest: the detection probability. 
The uncertainty relation found here can be extended to other observables. 
In \cite{Yin} a time-energy relation was discovered for the fluctuations
of the return time,  with an interesting dependence on the winding number
of the problem. 

As for possible experimental observation, these have demonstrated already the
quantum walk using single neutral
particles and site resolved microscopy  \cite{Karski,Bloch}. 
Usually the focus is  on the measurement
of the propagation of the packet of particles. 
This demands what we may call a global measurement searching for
the position of the particle at time $t$, while we are considering a spatially
local measurement which detects the particle on a node of a graph. 
Such experiments, on the recurrence problem, were conducted in \cite{Nitsche} 
 with coherent light
using strong projections, the number of repeated measurements was roughly forty.   
Thus  measurements of the quantum  detection probability $P_{{\rm det}}$ and the uncertainty principle,
are  within reach.  

 Usual uncertainty relations
are  statements showing the departure of quantum reality from classical Newtonian mechanics. While here we are dealing with the departure of quantum
search from its classical random walk counterpart.
 In Eq. (\ref{eqRob})  
we use the fluctuations of energy in the detected state
 $\mbox{Var}(H)_d$ and it is natural to wonder for its 
meaning in a measurement protocol. 
The observer repeatedly attempts to detect the
particle 
and once successful, namely the particle is detected, 
the particle is in state $\ket{{\rm d}}$. 
 Now that the particle
is detected we stop the monitoring  measurements on $\ket{{\rm d}}$. This means that
in this second stage of the experiment the energy
is a constant of motion.  
We now measure $H$. 
 Hence repeating the protocol   many times we have
from the first stage of the experiment
 an estimate for $P_{{\rm det}}$ and from the second the  variance of $H$ in the detected
state is obtained. 
It follows that at least in principle there is a  physical meaning to the variance of
$H$ in the detected state, as these are the fluctuations
after the particle is finally detected. 
It follows that we may rewrite Eq. 
(\ref{eqRob}) in a form that emphasizes the rule of the state function.
Since the final wave function, after a successful detection is 
$\ket{\psi_{{\rm fin}}} = \ket{{\rm d}}$ we have 
\begin{equation}
\Delta P \mbox{ Var}(H)_{{\rm \psi_{{\rm fin}}}}  \ge | \bra{\psi_{{\rm fin}} } \left[ H, D  \right] \ket{\psi_{{\rm in}}}|^2.
\label{eqnewRob}
\end{equation}
 The same holds more generally for $s \neq 1$. Note
that $\mbox{Var}(H)_{\rm \psi_{{\rm fin}}}$ is a constant of motion
after the  successful detection,  since as mentioned
 we stop the repeated detection
attempts  once obtaining the yes click.
This means that the observer  does not need to measure the fluctuations
immediately after the successful  detection and  there is no issue with
the  violation of energy time principle. Of course in practice one will have time
limitations since once decoherence kicks in,  due for example  to coupling to an environment,
the idealised model  considered here demands modifications.

{\bf Acknowledgement:} 
The support of Israel Science Foundation's grant 1898/17 is acknowledged.
FT is supported by DFG (Germany) under grant TH $2192/1-1$

\appendix
\section{}

 We present 
 calculations the single dangling bond system. 
The localised states describing the nodes of the
graph   $\ket{0}, \cdots\ket{6}$,  we measure on
$\ket{0}$ and 
we start on any other  node (in the text we also considered a uniform initial
condition). 
 The Hamiltonian is given by
$$
H = \left(
\begin{array}{c c c c c c c} 
0 & 1 & 0 & 0 & 0 & 0 & 0 \\
1 & 0 & 1 & 0 & 0 & 0 & 0 \\
0 & 1 & 0 & 1 & 0 & 0 & 1 \\
0 & 0 & 1 & 0 & 1 & 0 & 0 \\
0 & 0 & 0 & 1 & 0 & 1 & 0 \\
0 & 0 & 0 & 0 & 1 & 0 & 0 \\
0 & 0 & 1 & 0 & 0 & 0 & 0 
\end{array}
\right).
$$
$$ $$
From $H\ket{{\rm d}}, ... H^k \ket{{\rm d}} \cdots$ we get six normalised bright states
$$
\ket{0}, \ket{1} , {\ket{0} + \ket{2} \over \sqrt{2}},
{ 2 \ket{1} + \ket{3} + \ket{6} \over \sqrt{6}} ,$$
$$ { 2 \ket{0} + 4 \ket{2} + \ket{4} \over \sqrt{21} },
{ 6 \ket{1} + 5 \ket{3} + \ket{5} + 4 \ket{6} \over \sqrt{78}} . $$
Here $H^7 \ket{{\rm d}}$ yields a state which is a linear combination
of these states so it is of course excluded from this list. 
Since we have the dimension of the Hilbert state
equal  seven, we have one dark state orthogonal to the bright
ones. This is  easy to find $\ket{\delta} = (\ket{6} -\ket{3} + \ket{5})/\sqrt{3}$.
and we notice that this is a stationary state of the system so 
$\mbox{Var}(H)_{\delta}  = 0$.  It follows that the detection probability
$P_{{\rm det}}$ 
is equal   unity for the transitions $0, 1,  2, 4 \rightarrow 0$
while for $3, 5, 6 \rightarrow 0$ we find $ P_{{\rm det}} =
2/3$. This is presented
in the Fig.  The uncertainty principle gives the bound
$P_{{\rm det }} (1\to 0) \ge 1, 
P_{{\rm det }} (2\to 0) \ge 1, 
P_{{\rm det }} (3\to 0) \ge 1/6, 
P_{{\rm det }} (4\to 0) \ge 1/17, 
P_{{\rm det }} (5\to 0) \ge 1/78$,
      see Fig. \ref{figItai}.
For the $5 \to 0$ transition we need $\mbox{Var}(H^5)_0 = 78$ 
which is larger than the fluctuations of energy when $s=1,..4$.

%\begin{figure}
%\centering
%\includegraphics[width=0.99\columnwidth]{itayEX.pdf}
%\caption{
%Itay's examples      }
%      \label{figItaiExamples}
%    \end{figure}


\begin{thebibliography}{99}

%Uber eine Aufgabe der Wahrscheinlichkeitsrechnung betreffend die Irrfahrt im Strassennetz ¨ ,
\bibitem{Polya} Georg P\'olya 
{\em Math. Ann.} {\bf  84} 149  (1921).
% 149-160.

\bibitem{Redner} S. Redner,
{\em A Guide to First-Passage Processes}
Cambridge University Press (2007).

\bibitem{Ralf} R. Metzler, G. Oshanin, and S. Redner {\em First-Passage Phenomena and their applications}
World Scientific  (2014).

% quantum random walks
\bibitem{Aharonov1}
Y. Aharonov, L. Davidovich, and N. Zagury
{\em Phys. Rev. A.} {\bf 48}, 1687 (1993).

% One-dimensional quantum walks
\bibitem{Ambainis}
A. Ambainis, E. Bach, A. Nayak, A. Vishwanath,  A. Watrous
Proceeding
STOC $'01$ Proceedings of the thirty-third annual ACM symposium on Theory of computing
Pages $37-49$ (2001).  

% Continuous time quantum walks: Models for coherent transport on
% complex networks
\bibitem{Blumen} O. M\"ulken, and A. Blumen {\em Phys. Rep.} {\bf 502}
37 (2011).

\bibitem{Salvador} S. E. Venegas-Andraca
{\em Quantum Information Processing}  {\bf 11}(5), 1015 (2012)





% One dimensionla quantum walks with absorbing
\bibitem{Bach}
E. Bach, S. Coppersmith, M. P. Goldschen,
R. Joynt,  J. Watrous
{\em J. of Computer and System Science}
{\bf 69} 562 (2004).

% Hitting time for quantum walks on the hypercube
% Symmetry plays an important role in the dramatic speed ups
%and slow dows of quantum walks.
% Here the cube is discussed
\bibitem{Krovi} H. Krovi, and T. A.  Brun
{\em Phys. Rev. A} {\bf 73}, 032341 (2006).


% Quantum walks with infinite hitting time
\bibitem{Krovi1}
H. Krovi, and T. A. Brun
{\em Phys. Rev. A} 74, 042334  (2008).

% Hitting time for the continuous quantum walk
\bibitem{Varbanov}
M. Varbanov, H. Krovi, and T. A. Brun,
{\em Phys. Rev. A} {\bf 78}, 022324 (2008). 


%{\em  Recurrence for discrete time unitary evolutions}
\bibitem{Grunbaum} F. A. Gr\"unbaum, L. Vel\'azquez, A. H. Werner and R. F. Werner,
{\em Comm. Math. Phys.}, {\bf 320} 543 (2013).

% Survival of classical and quantum particle in the presence of traps
\bibitem{Krapivsky}
P. L. Krapivsky, J. M. Luck, and K. Mallick
{\em J. Stat. Phys.} {\bf  154} 1430 (2014). 

% Non-Hermitian
%{\em  Detection of a quantum particle on a lattice and repeated projective measurements}
\bibitem{Dhar} S. Dhar, S. Dasgupta, A. Dhar, and D. Sen
{\em Phys. Rev. A.} {\bf 91}, 062115 (2015).


%{\em  Quantum time of arrival distribution in a simple lattice model}
\bibitem{Dhar1}
S. Dhar, S. Dasgupta, and A. Dhar,
{\em J. Phys. A} {\bf 48}, 115304 (2015).

\bibitem{Sinkovicz1} P. Sinkovicz, T. Kiss, and J. K. Asboth
{\em Phys. Rev. A.} {\bf 93}, 050101(R) (2016).

%{\em Quantum walks: the first detected passage time problem}
\bibitem{Harel}
H. Friedman, D. Kessler, and E. Barkai
{\em Phys. Rev. E.}
{\bf 95}, 032141 (2017).

 

%{\em First detected arrival of a quantum walker on an infinite line}
\bibitem{Felix1}
F. Thiel, E. Barkai, and D. A. Kessler
{\em Phys. Rev. Lett.} {\bf 120}, 040502 (2018).

% {\em Spectral dimension controlling
%the decay of the quantum first-detection probability}
\bibitem{FelixPRA}
F. Thiel, D.A. Kessler and E. Barkai,
{\em Phys. Rev. A}
{\bf 97}, 0621015 (2018).

%Return to the origin problem for a particle on a one-dimensional lattice with quasi-Zeno dynamics
\bibitem{Lahiri}
S. Lahiri, and A. Dhar
{\em Phys. Rev. A} 
{\bf 99}. 012101 (2019).

%Quantum Dynamics under continuous projective measurements: non-Hermitian description and the continuous space limit
\bibitem{Dubey} V. Dubey, C. Barnardin, A. Dhar
	arXiv:2012.01196 [quant-ph]
(2020). 

%{\em Quantum walks: the mean first detected transition time}
\bibitem{Liu}
Q. Liu, R. Yin, K. Ziegler, and E. Barkai
{\em Physical Review Research}
{\bf 2}, 033113 (2020).



\bibitem{Caruso} F. Caruso, A. W. Chin, A. Datta, S. F. Huelga, and M. B. Plenio
 {\em J. Chem. Phys.} {\bf 131}, 105106 (2009).

%Uncertainty and symmetry bounds for the total detection probability of quantum
%walks}
\bibitem{FelixLET} F. Thiel, I. Mualem,  D. Kessler, and E. Barkai  
F. Thiel, I. Mualem,  D. Kessler and E. Barkai
{\em Physical Review Research}
2, 023392, (2020).

%arXiv:1906.08108 [quant-ph] (2019). 

% Dark states of quantum search cause imperfect detection
\bibitem{FelixExactKB} 
F. Thiel, I. Mualem, D. Meidan,  E. Barkai, and D. Kessler
{\em Physical Review A}
{\bf 102}, 02210 (2020).
%https://arxiv.org/abs/1912.08649




%{\em Large fluctuations of the first detected quantum return time}
\bibitem{Yin} 
R. Yin, K. Ziegler, F. Thiel, and E. Barkai
{\em Physical Review Research} {\bf 1}, 033086 (2019).

% 
\bibitem{Busch}
P. Busch,  T. Heinonen, and P. Lahti
{\em Phys. Rep.} {\bf 452}, 155 (2007).


% Quantum Zeno Subspaces and Decoherence
% Int. Sympo. on Fundamental Physi
\bibitem{Facchi} P. Facchi, S. Pascazio
{\em J. Phys. Soc. Japan} {\bf 73} Suppl. C 30-33 (2003).

% Quantum Zeno dynamics: mathematical and physical aspects
\bibitem{Pascazio} P. Facchi, S. Pascazio {\em J. Phys. A: Math. Theor.}
{\bf 41} (2008).


% The scope of Zeno protocols is to constrain the dynamics of the
% system to remain within a sub-space of the Hilbert space, which is
% thus called Zeno subspace. This can be accomplished when
%
% Quantum  Zeno Dynamics Through Stochastic Protocols
\bibitem{Gherardini} M. M. M\"uller, S. Gherardini, and F. Caruso
{\em annalen der Physik} {\bf 529} 1600206 (2017).


% "The Zeno's paradox in quantum theory". 
\bibitem{ZenoIN} 
B. Misra, 
and E. C. G. Sudarshan
{\em Journal of Mathematical Physics} {\bf 18} 756  (1997). 

%Violation of Heisenberg’s Measurement-Disturbance Relationship by Weak Measurements
\bibitem{Rozema}
L. A. Rozema, et al. 
{\em Phys. Rev. Lett.}  {\bf 109}, 100404 (2012).


\bibitem{Heisenberg} W. Heisenberg, {\em Z. Phys.} {\bf 43}, 172 (1927). 

\bibitem{Kenard}
E.H. Kennard, {\em Z. Phys.} 44, 326 (1927).

%The uncertainty principle. 
\bibitem{Robertson}
H. P. Robertson,
{\em Phys. Rev.} {\bf 34}, 163–164 (1929).

\bibitem{Vlad}
V. B. Braginsky, and F. Y. Khalili
{\em Qunatum Measurement} Cambridge Press (1992). 

% Quantum speed limits: from Heisenberg's uncertainty principle
%to optimal quantum control
\bibitem{Deffner}
S. Deffner, and S. Campbell
{\em J. Phys. A: Math. Theor.}   {\bf 50}, 453001 (2017)

\bibitem{Ozawa}
M. Ozawa,  
{\em Phys. Rev. A}  {\bf 67}, 042105 (2003).
ibid. {\em Ann. Phys.} {\bf 311}, 350-416 (2004).

%Universally valid reformulation of the Heisenberg uncertainty principle on noise and disturbance in measurements. 
% Experimental demonstration of a universally valid error-disturbance uncertainty relation in spin measurements
\bibitem{Erhart}
J. Erhart et al. {\em Nature Physics.} {\bf 8} 185 (2012). 

%The uncertainty principle determines the non-locality of quantum mechanics 
\bibitem{Oppenheim} 
J. Oppenheim, and  S. Wehner, 
{\em Science} {\bf  330}, 1072–1074 (2010).

% Proof of heisenberg's error disturbance relation
\bibitem{Werner}
P. Busch,  P. Lahti, and R. F. Werner
{\em Phys. Rev. Lett.} 111, 160405 (2013). 



%"Stronger uncertainty relations for all incompatible observables", 
\bibitem{STRONGER}
L. Maccone and A. K. Pati, 
{\em Phys. Rev. Lett.} {\bf  113}, 260401 (2014).

% Relativistic independence bounds nonlocality
\bibitem{ECOHEN}
A. Carmi, and E. Cohen
{\em Science Advances} {\bf 5} no. 4 eaav8370 (2019).



\bibitem{Karski}
 M. Karski,  et al.  
{\em Science} {\bf  325} 174 (2009).

\bibitem{Bloch} J. F. Sherson, et al.
{\em Nature}
{\bf  467}  68
(2010). 

% Probing Measurement Induced Effects in Quantum Walks via Recurrence
\bibitem{Nitsche}
T. Nitsche. et al {\em Science Advances} {\bf 4} eaar6444 (2018). 






\end{thebibliography}
\end{document}